\documentclass{PoS}
\newcommand{\asmz}{\alpha_s(M_Z^2)}
\newcommand{\msbar}{\mbox{$\overline{\rm{MS}}$}\ }
\title{HERAPDF2.0Jets NNLO (prel.), the completion of the HERAPDF2.0 family}

\ShortTitle{HERAPDF2.0Jets NNLO}

\author{\speaker{A.~M.~Cooper-Sarkar}\thanks{On behalf of the H1 and ZEUS collaborations}\\
        Unniversity of Oxford\\
        E-mail: \email{amanda.cooper-sarkar@physics.ox.ac.uk}}

\abstract{The HERAPDF2.0 family, introduced in 2015, is completed
  with fits HERAPDF2.0Jets NNLO (prel.) based on inclusive
  HERA data and selected jet production data.
  The result of a fit with the strong coupling constant, $\asmz$, free is
  $\asmz = 0.1150 \pm 0.0008{\rm (exp)} ^{+0.0002}_{-0.0005}$ 
  (model/parameterisation) $\pm 0.0006{\rm (hadronisation)} \pm 0.0027 {\rm (scale)}$.
  Sets of parton density functions, PDFs, from fits with fixed
  $\asmz = 0.115$ and  $\asmz = 0.118$ are presented and compared.}

\FullConference{XXVII International Workshop on Deep-Inelastic Scattering and Related Subjects - DIS2019\\
		8-12 April, 2019\\
		Torino, Italy}

\begin{document}
\section{Introduction \label{sec:int}}

Deep inelastic scattering (DIS) of electrons
on protons, $ep$, at centre-of-mass energies of up to $\sqrt{s} \approx 320\,$GeV
at HERA has been central to the exploration
of proton structure and quark--gluon dynamics as
described by perturbative Quantum Chromo Dynamics (pQCD). 
The combination of H1 and ZEUS data on inclusive $ep$ scattering
and the subsequent pQCD analysis, introducing the family of 
parton density functions (PDFs) known as HERAPDF2.0\cite{HERAPDF20},
was a milestone for the exploitation of the HERA data.
The preliminary work presented here represents a completion of the 
HERAPDF2.0 family with a fit at NNLO to
HERA inclusive and jet production data
published separately by the
ZEUS and H1 collaborations.
This was not possible at the time of the original introduction of HERAPDF2.0
because a treatment at NNLO of jet production in 
$ep$ scattering was not available then.

\section{Procedure and Data}
The name HERAPDF stands for a pQCD analysis within the
DGLAP formalism, where predictions from pQCD are fitted to data.
These predictions
are obtained by solving the DGLAP evolution 
equations at LO, NLO and NNLO in the \msbar scheme. 
The inclusive and dijet production data which were already
used for HERAPDF2.0Jets NLO were again used for the analysis presented here.
A new data set~\cite{h1lowq2newjets} published by the 
H1~collaboration on jet production in low~$Q^2$ events, 
where $Q^2$ is the four-momentum-transfer squared,
was added as input to the fits.

The fits presented here were done in the same way 
as for all other members of the HERAPDF2.0 family, for full details see~\cite{h1zeusprelim} and references therein.
The fits were performed using the programme 
QCDNUM within the xFitter framework. 
Only cross sections for $Q^2$ starting at $Q^2_{min} = 3.5$\,GeV$^2$ 
were used in the analysis. 
All parameter setting were the same as for the HERAPDF2.0Jets NLO fit.
The analysis of uncertainties was also performed in exactly the same way.

There were some modifications with respect to the analysis at NLO.
They were driven by the usage of
the newly available treatment of jet production at NNLO.
The jet data were included in the fits at NNLO using
predictions for the jet cross sections calculated using NNLOJET~\cite{nnlojet},
which was interfaced to the fast interpolation grid code, fastNLO and
Applgrid using the Applfast framework~\cite{applfast} 
in order to achieve the required speed for the convolution for us in an iterative PDF fit.
The predictions were multiplied by corrections for hadronisation and 
$Z^0$ exchange before they were used in the fits.
A running electro-magnetic $\alpha$ as implemented in the 2012 version of 
the programme EPRC was used for the treatment
of the jet cross sections.

The new treatment of inclusive jet and dijet production at NNLO was only 
applicable to a slightly reduced phase space compared to 
HERAPDF2.0Jets NLO. All data points 
with $\sqrt{\langle p_T^2 \rangle +Q^2} \le 13.5$\,GeV were excluded,
where $p_T$ is the transverse energy of the jets. 
In addition, six data points, the lowest $\langle p_T \rangle$ 
bin for each $Q^2$ region, were excluded from the ZEUS dijet
data set because the NNLO  predictions for these points were
deemed unreliable.
In addition, the trijet data which were used as input
to HERAPDF2.0Jets NLO had to be excluded as their treatment at NNLO was not available.

The choice of scales was also adjusted for the NNLO analysis.
At NLO, the factorisation scale was chosen as 
$\mu_{\rm f}^2 = Q^2$,
while the renormalisation scale was linked to the transverse
momenta, $p_T$, of the jets by $\mu_{\rm r}^2 = (Q^2 + p_{T}^2)/2$.
For the NNLO analysis, $\mu_{\rm f}^2 =\mu_{\rm r}^2= Q^2 + p_{T}^2$
was chosen.

\section{Determination of the strong coupling constant}
\label{sec:as}

Jet production data are essential for the determination of
the strong coupling constant, $\asmz$.
In pQCD fits to inclusive DIS data alone, the gluon PDF is determined 
via the DGLAP equations only, using the observed scaling violations.
This results in a strong correlation between the shape of the 
gluon distribution and the value of $\asmz$. 
Data on jet production cross sections provide an independent constraint
on the gluon distribution.
Jet and dijet production are also directly sensitive to $\asmz$ and 
thus  such data allow for an accurate simultaneous determination of $\asmz$
and the gluon distribution.

The HERAPDF2.0Jets NNLO (prel.) fit with free $\asmz$
gave a value of
\begin{eqnarray}
\nonumber
\asmz =0.1150 \pm 0.0008{\rm (exp)} ^{+0.0002}_{-0.0005}{\rm (model/parameterisation)} \\
\nonumber
 ~~~~ \pm 0.0006{\rm (hadronisation)}~~ \pm 0.0027 {\rm (scale)}~~.
\end{eqnarray}
This result on $\asmz$ is compatible with the world average~\cite{PDG18} 
and it is competitive with other determinations at NNLO.

The HERAPDF2.0Jets NNLO (prel.) fit with free $\asmz$  uses
1343 data points and has a
$\chi^2/$d.o.f.\,$= 1599/1328 = 1.203$. This can be compared
to the $\chi^2/$d.o.f.\,$= 1363/1131 = 1.205$ for HERAPDF2.0 NNLO based on
inclusive data only.
The similarity of the $\chi^2/$d.o.f. values indicates that 
the data on jet production do not introduce any tension.

\begin{figure}
  \centering
  \setlength{\unitlength}{0.1\textwidth}
%  \begin{picture} (9,12)
  \begin{picture} (9,8.5)
  \put(1.2,4.5){\includegraphics[width=0.60\textwidth]{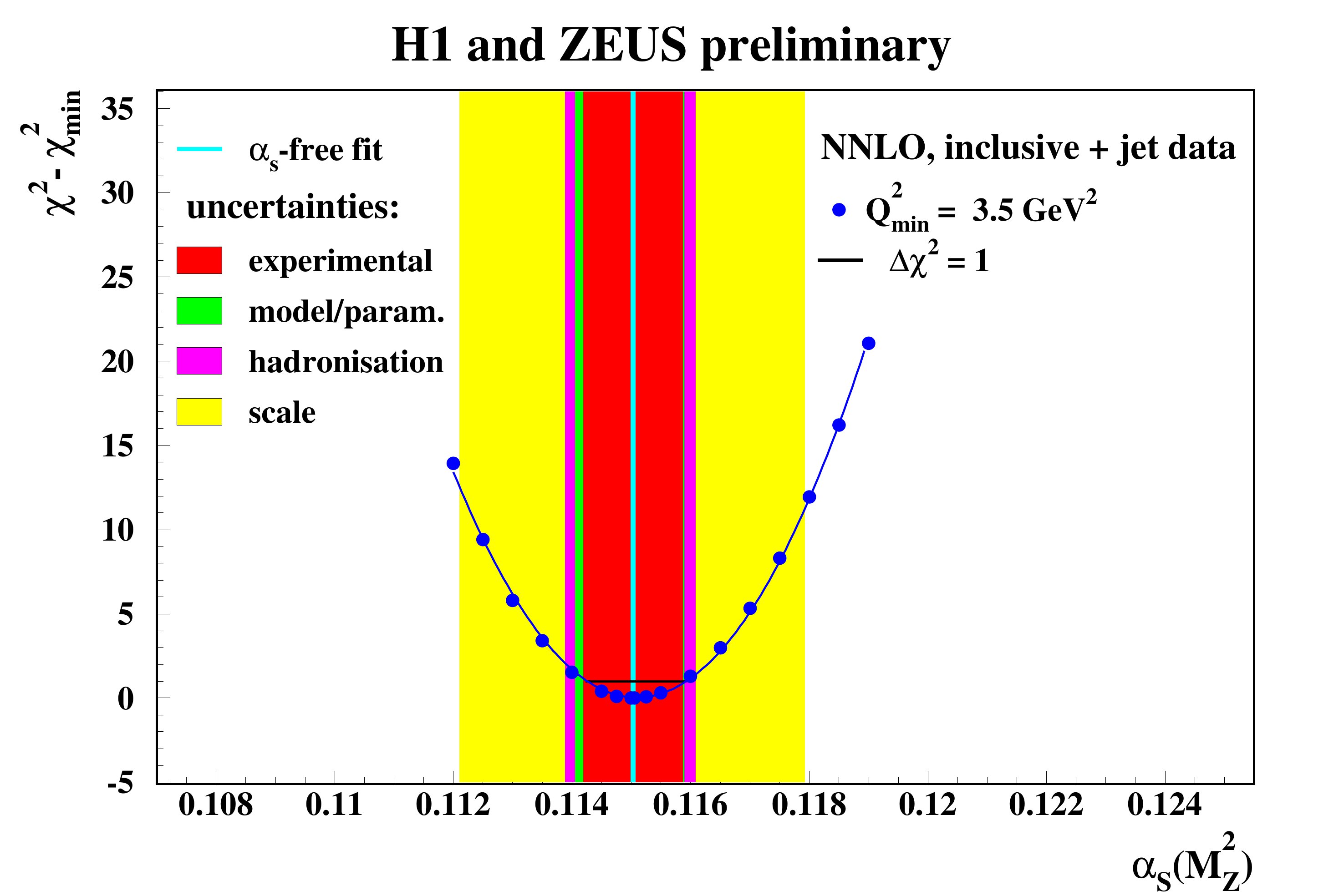}}
  \put(1.2,0.0){\includegraphics[width=0.65\textwidth]{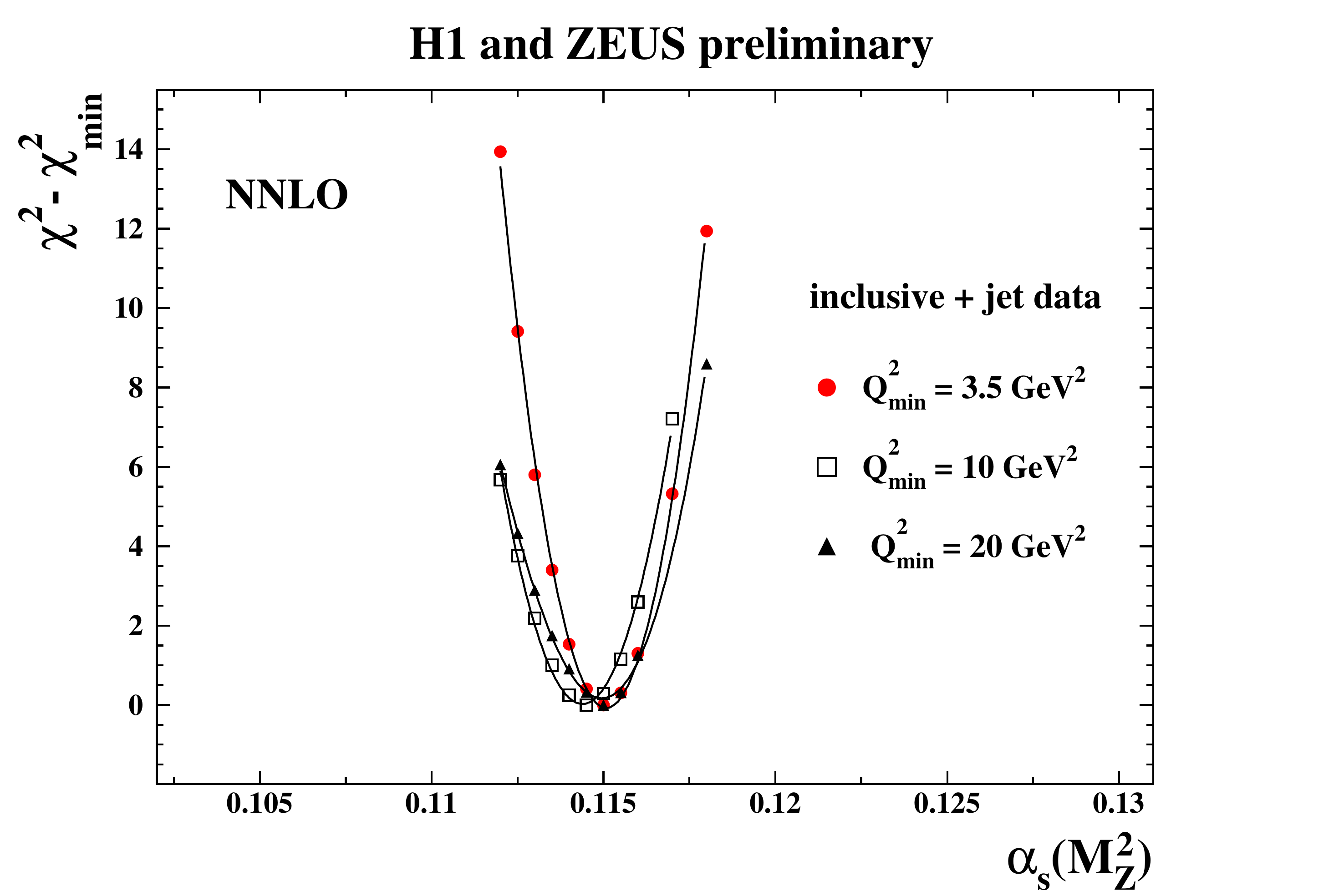}}
  \put (0.6,5.4) {a)}
  \put (0.6,0.9) {b)}
  \end{picture}
\caption {$\Delta \chi^2 = \chi^2 - \chi^2_{\rm min}$ vs.\ $\asmz$ for
HERAPDF2.0Jets NNLO (prel.) fits with fixed $\asmz$ with a) the standard
$Q^2_{min}$ of 3.5\,GeV$^2$ b) with $Q^2_{min}$ set to
3.5\,GeV$^2$, 10\,GeV$^2$ and 20\,GeV$^2$ for the inclusive data.
}
\label{fig:alphasscan}
\end{figure}

The experimental uncertainty was determined from the fit.
The $\chi^2$ scan in $\asmz$ shown in Fig.~\ref{fig:alphasscan}a) confirmed
the value of $\asmz$ and the experimental uncertainty.
In addition to this the HERAPDF procedure considers model and parameterisation 
uncertainties and, for jet data, hadronisation uncertainties are also 
considered, see~\cite{HERAPDF20} for details. These additional uncertainties 
are also shown in Fig.~\ref{fig:alphasscan}a).

A strong motivation to determine
$\asmz$ at NNLO was the hope to substantially reduce scale uncertainties.
This uncertainty
was evaluated by varying the renormalisation and factorisation 
scales by a factor of two,
both separately and simultaneously, 
and taking the maximal positive and negative deviations.
The uncertainties were assumed to be 
50\,\% correlated and 50\,\% uncorrelated
between bins and data sets.
The result is also shown in Fig.~\ref{fig:alphasscan}a).
The scale uncertainty still dominates the uncertainties.

As the input data were changed for the NNLO
analysis and the choice of scales were changed with respect to the NLO analysis,
a detailed comparison of scale uncertainties will be published after
the appropriate reanalysis of the data at NLO.
However, the present scale uncertainty, of $\pm 0.0027$ for the NNLO analysis, 
is significantly lower
than the $+0.0037,-0.0030$ previously observed for the HERAPDF2.0Jets
NLO analysis.
If the NNLO determination of $\asmz$ is performed with the old
choice of scales, the value of $\asmz$ is further reduced to 0.1135, 
well within scale uncertainties.

The question whether data with relatively low $Q^2$ bias the
determination of $\asmz$ arises in the context of the
HERA data anlysis for which low $Q^2$ is also low $x$. A treatment beyond DGLAP
 may be necessary because of higher-tiwst terms, $ln(1/x)$ terms or even parton saturation. To check for such bias
Figure~\ref{fig:alphasscan}b) shows scans with  
$Q^2_{min}$ set to 3.5\,GeV$^2$, 10\,GeV$^2$ and 20\,GeV$^2$
for the inclusive data. Clear minima are visible which coincide within
uncertainties.

\section{The PDFs of HERAPDF2.0Jets NNLO (prel.)}

\begin{figure}[tbp]
  \centering
  \setlength{\unitlength}{0.1\textwidth}
  \begin{picture} (11,8.5)
  \put(0.0,4.2){\includegraphics[width=0.55\textwidth]{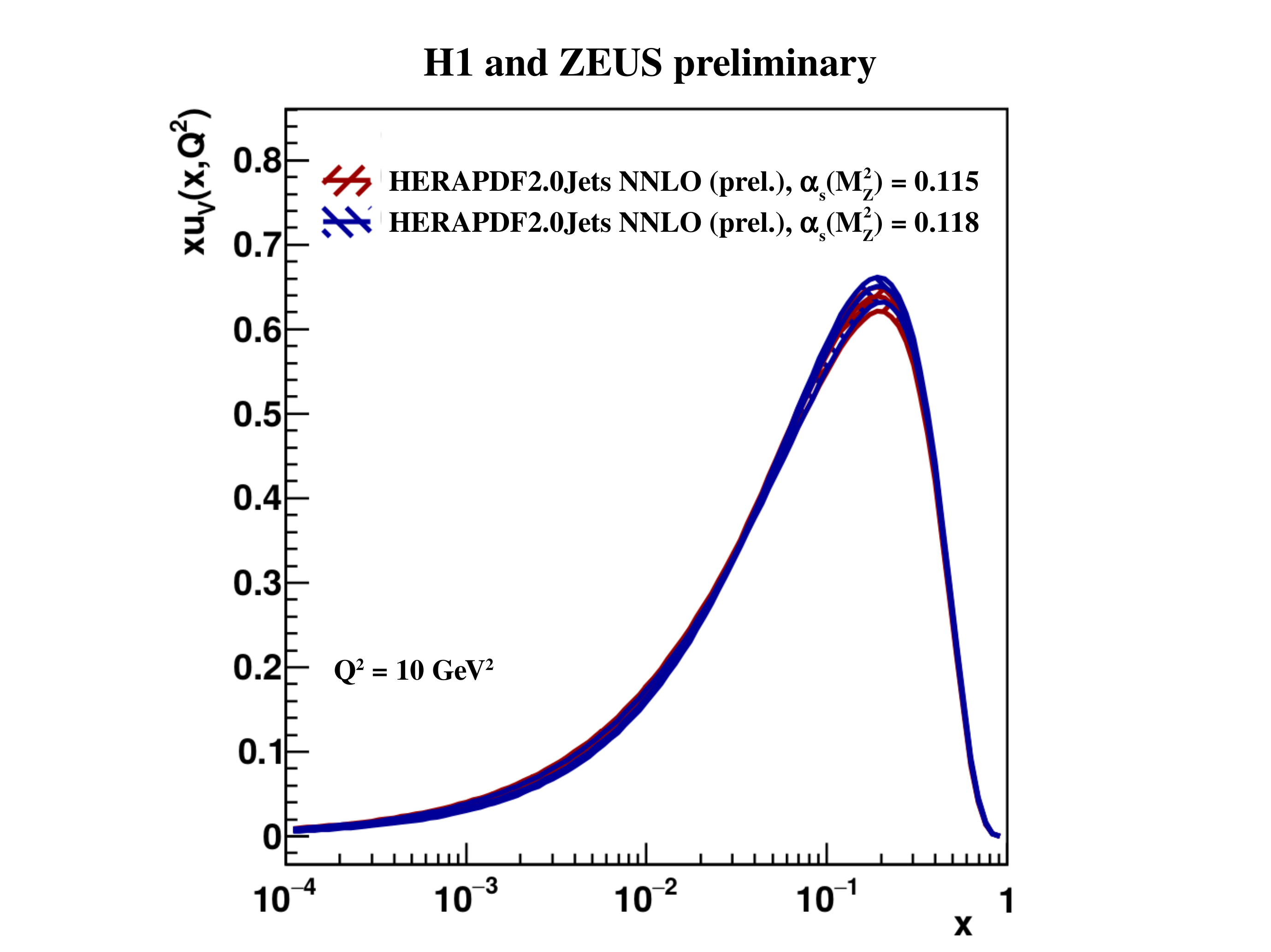}}
  \put(5.0,4.2){\includegraphics[width=0.55\textwidth]{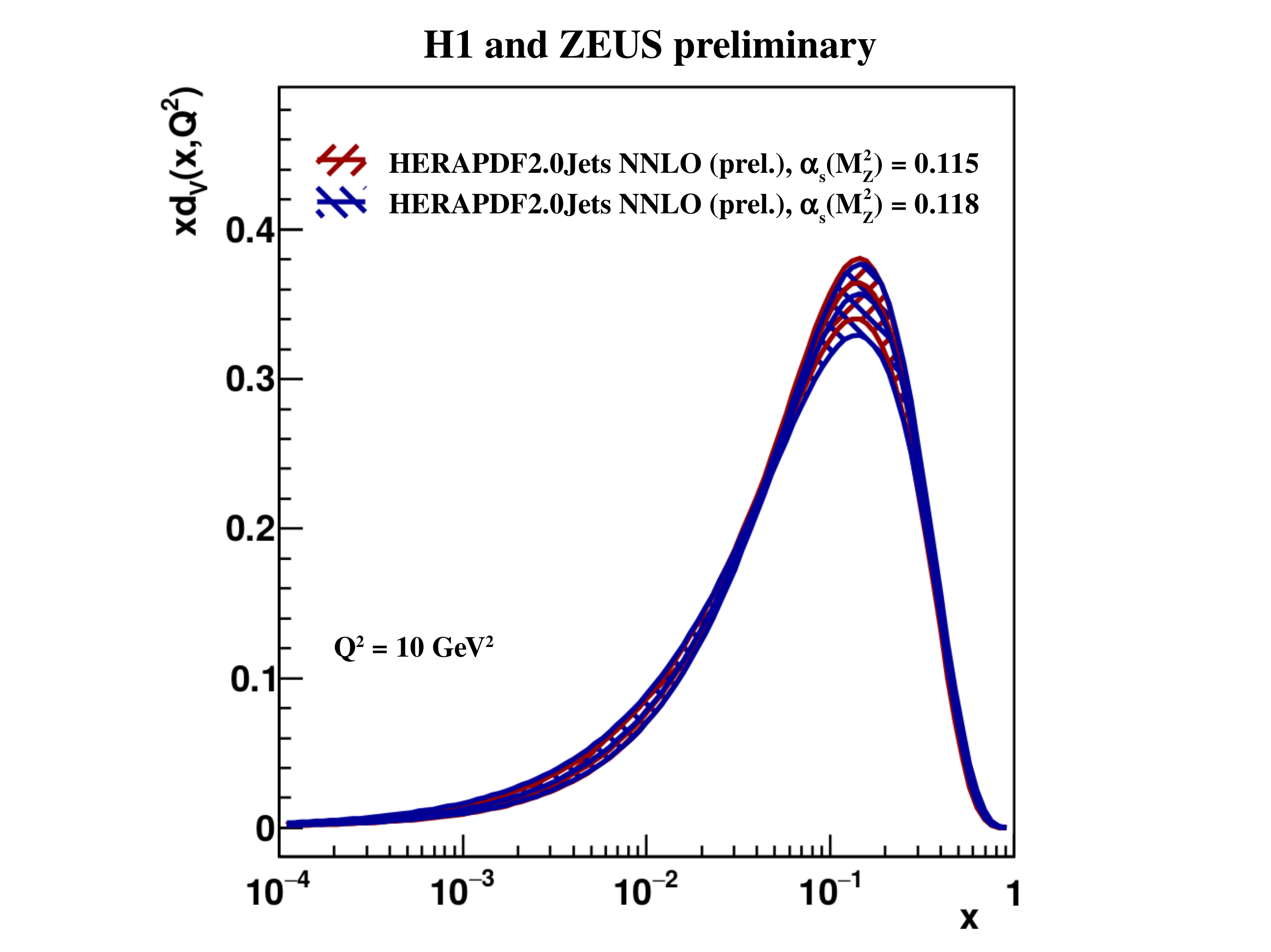}}
  \put(0.0,0.0){\includegraphics[width=0.55\textwidth]{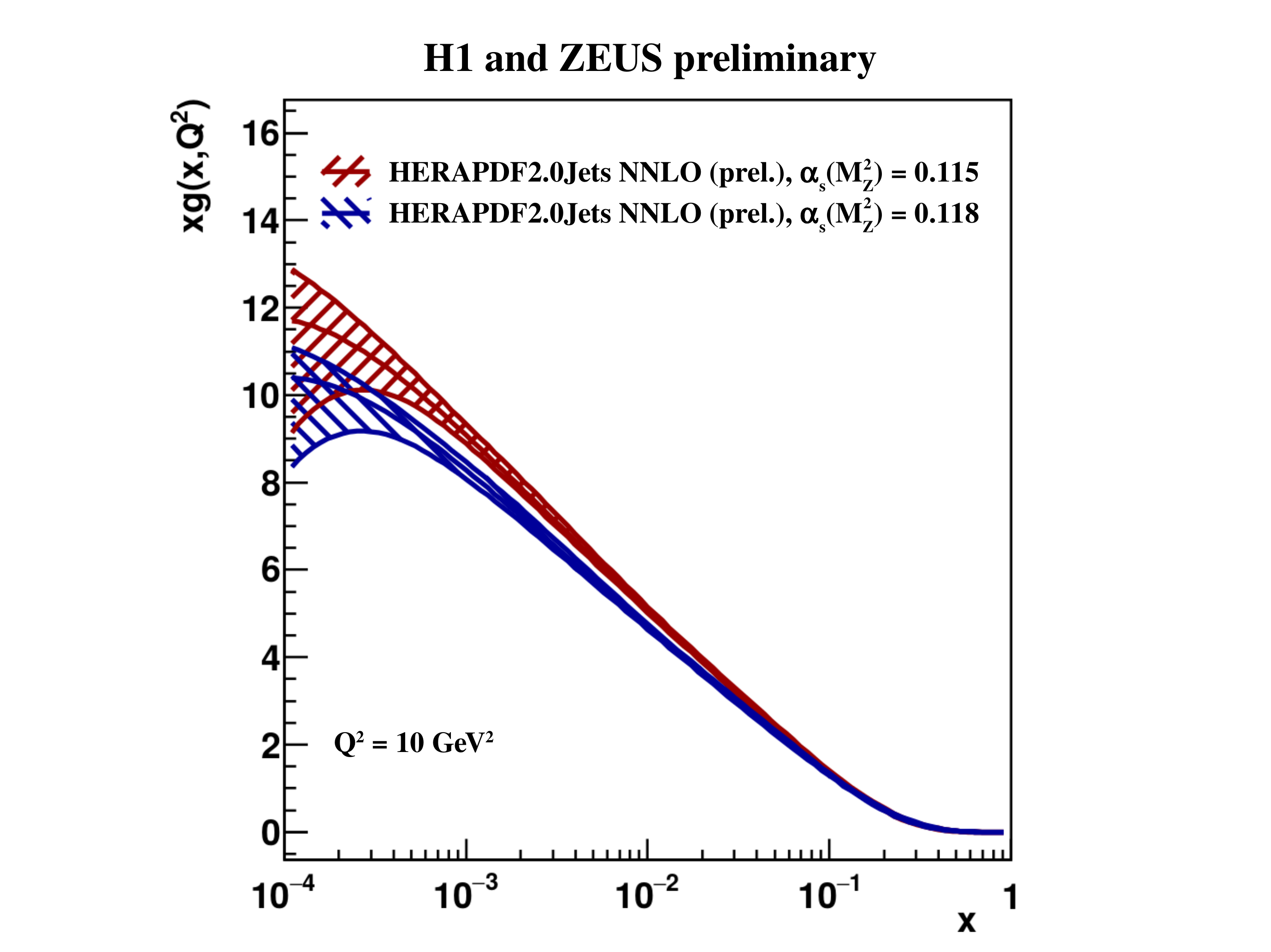}}
   \put(5.0,0.0){\includegraphics[width=0.55\textwidth]{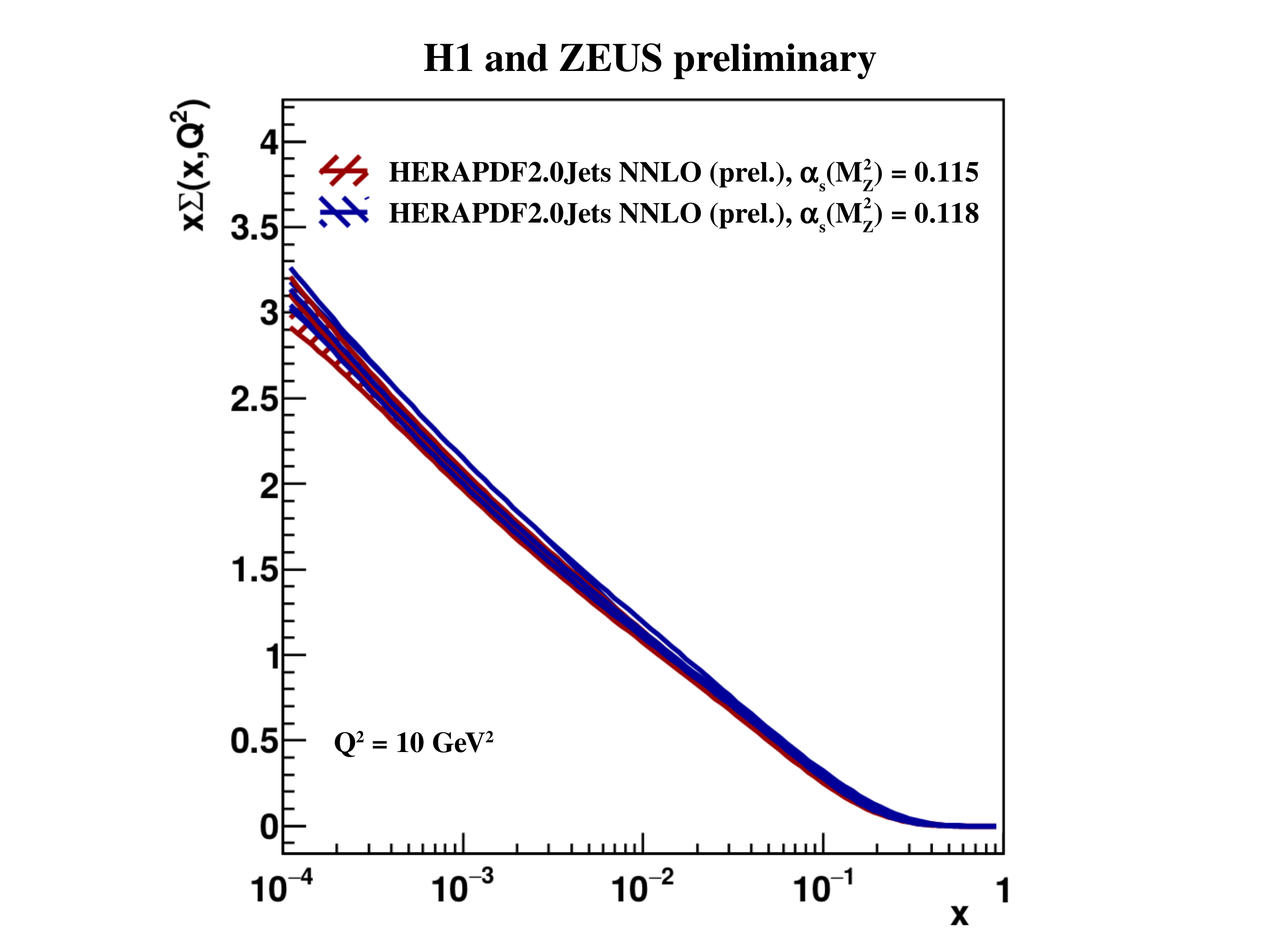}}
  \put (0.6,4.7) {a)}
  \put (5.6,4.7) {b)}
  \put (0.6,0.5) {c)}
  \put (5.6,0.5) {d)}
  \end{picture}
\vspace{-0.5cm} 
\caption { 
Comparison of the parton distribution functions 
a) $xu_v$, b) $xd_v$, c) $xg$ and d) $x\Sigma=x(\bar{U}+\bar{D})$ of 
HERAPDF2.0Jets NNLO (prel.) with fixed $\asmz = 0.115$ and $\asmz = 0.118$
at the scale $Q^{2} = 10\,$GeV$^{2}$.
The total uncertainties are shown as differently hatched bands.
}
\label{fig:as0-115vsas0-118}
\end{figure}
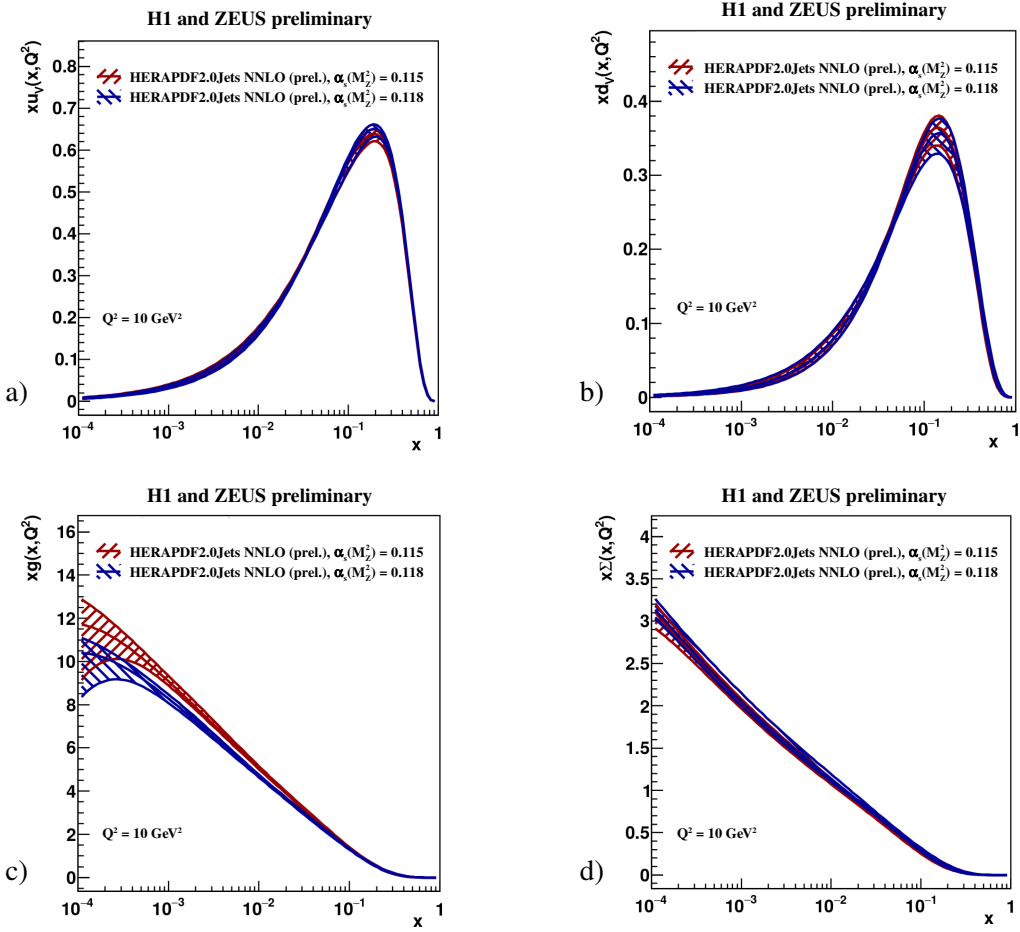

\begin{figure}[tbp]
  \centering
  \setlength{\unitlength}{0.1\textwidth}
  \begin{picture} (11,8.5)
  \put(0.0,4.2){\includegraphics[width=0.55\textwidth]{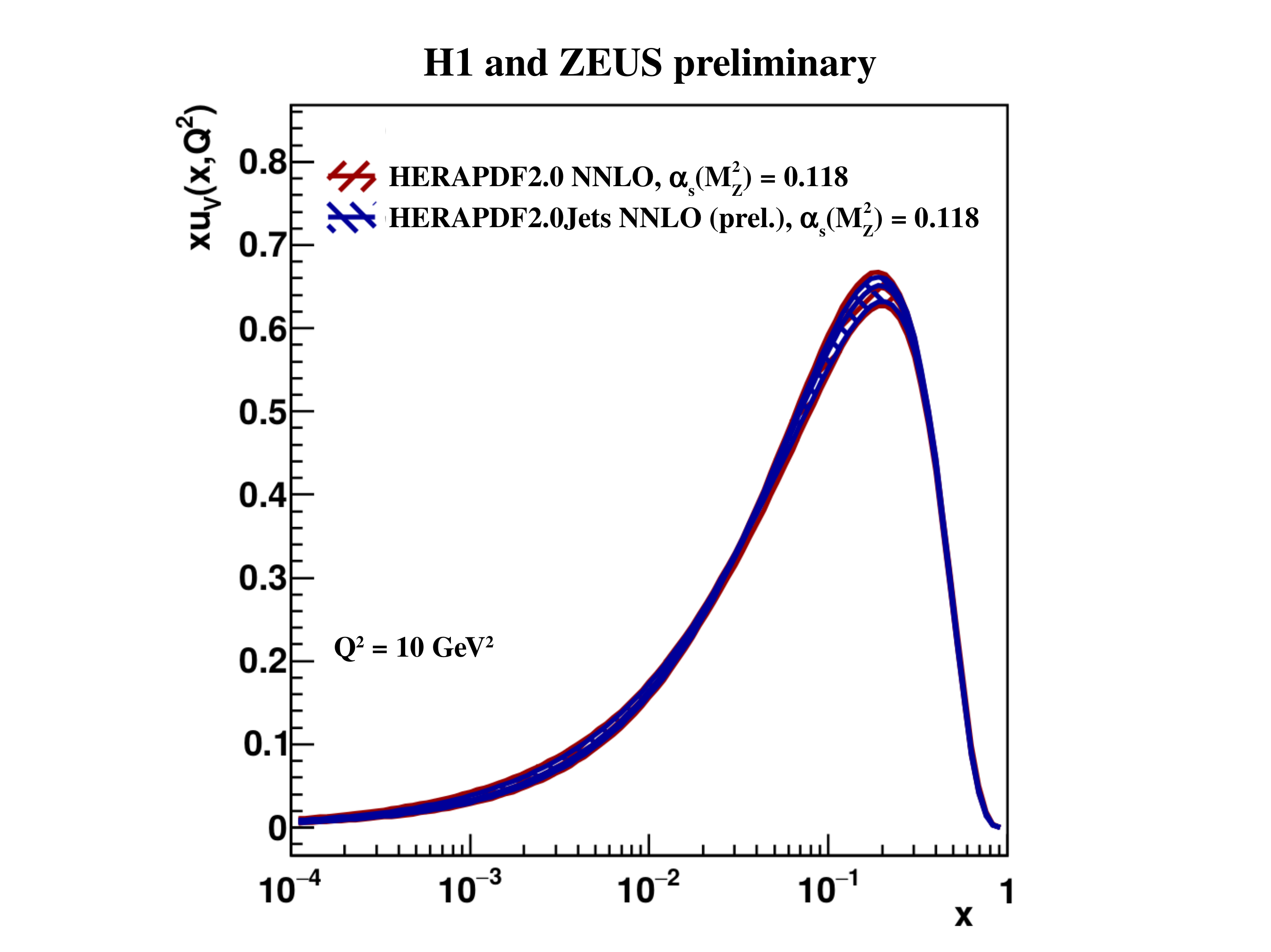}}
  \put(5.0,4.2){\includegraphics[width=0.55\textwidth]{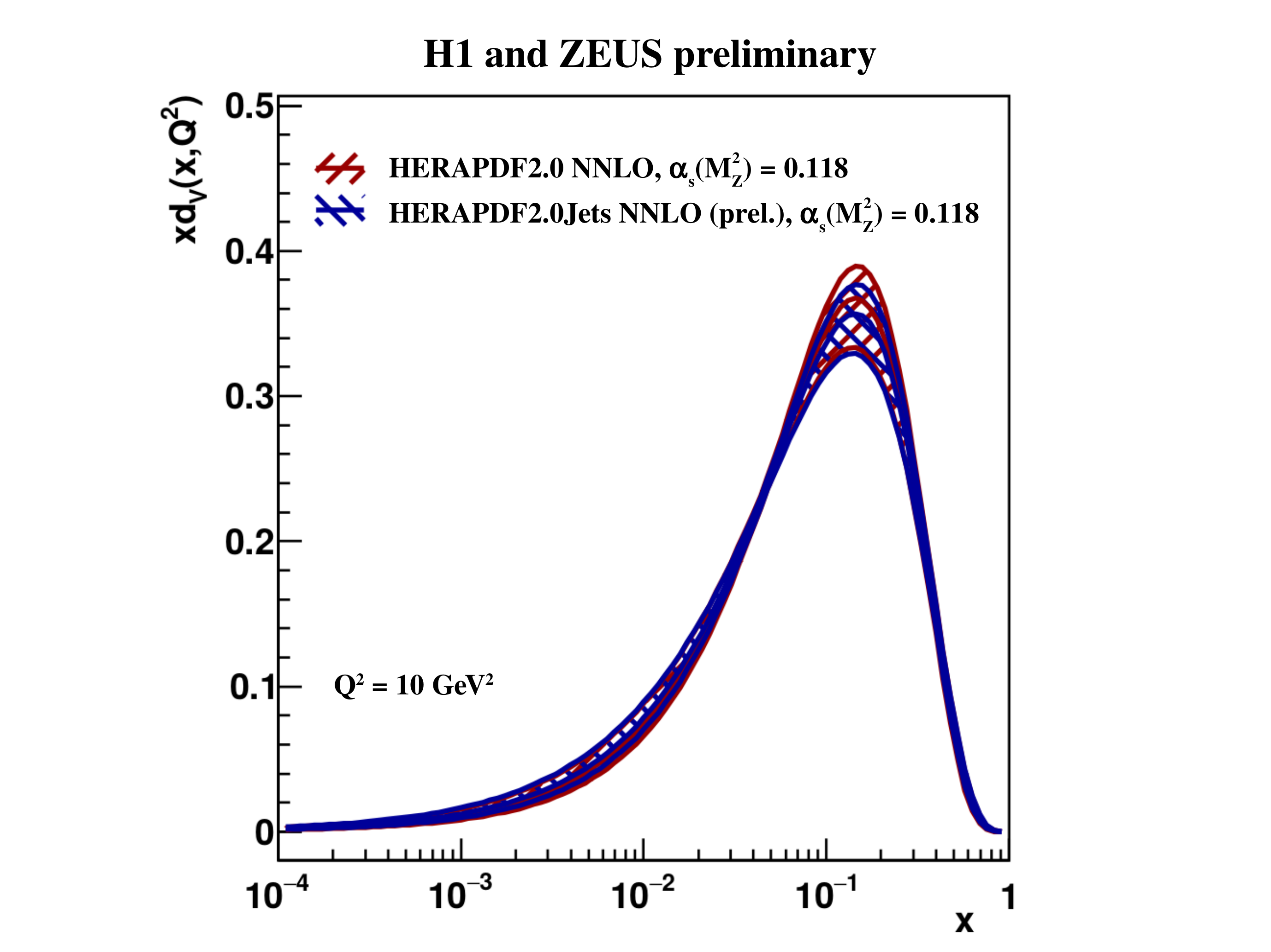}}
  \put(0.0,0.0){\includegraphics[width=0.55\textwidth]{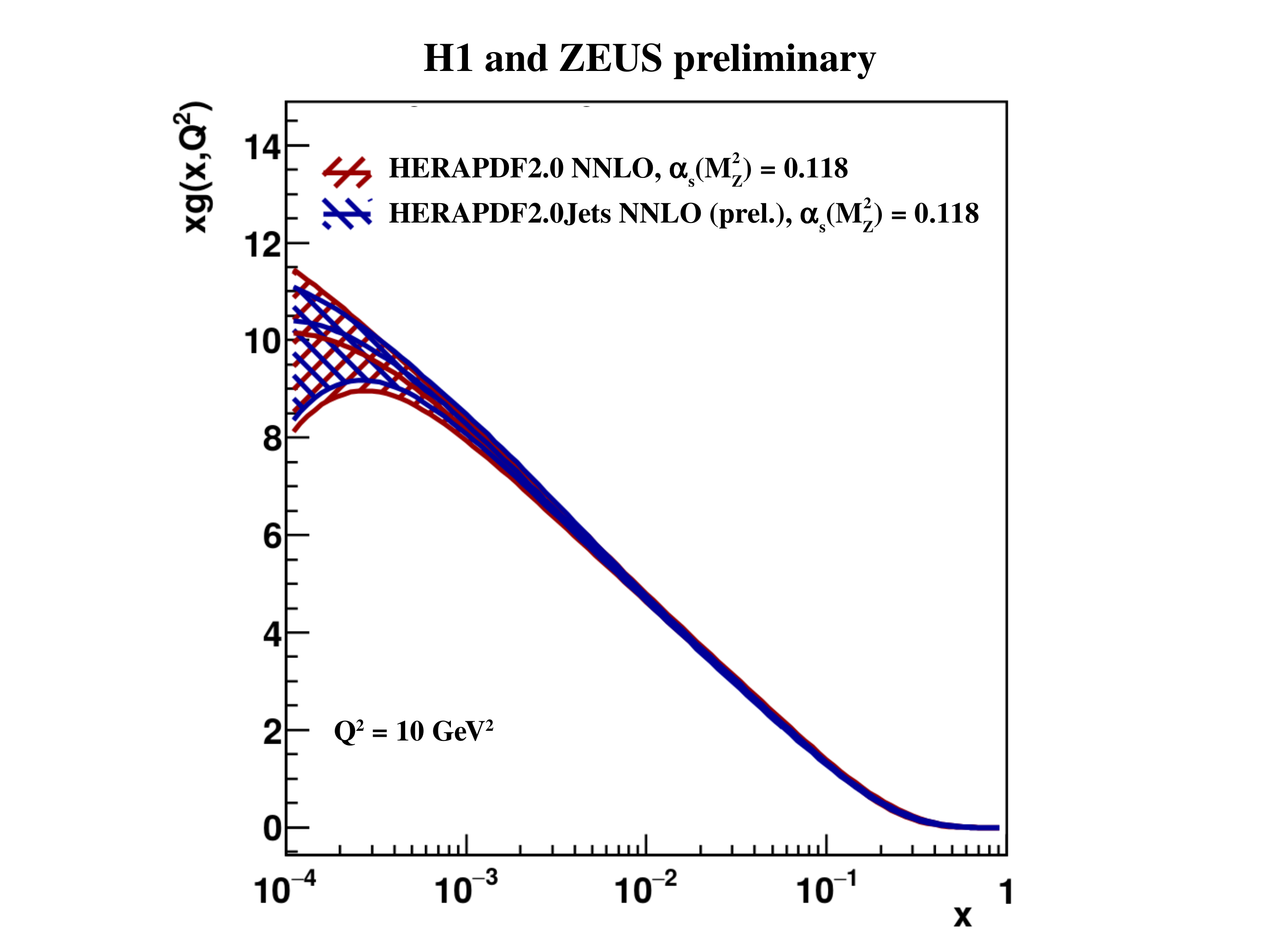}}
   \put(5.0,0.0){\includegraphics[width=0.55\textwidth]{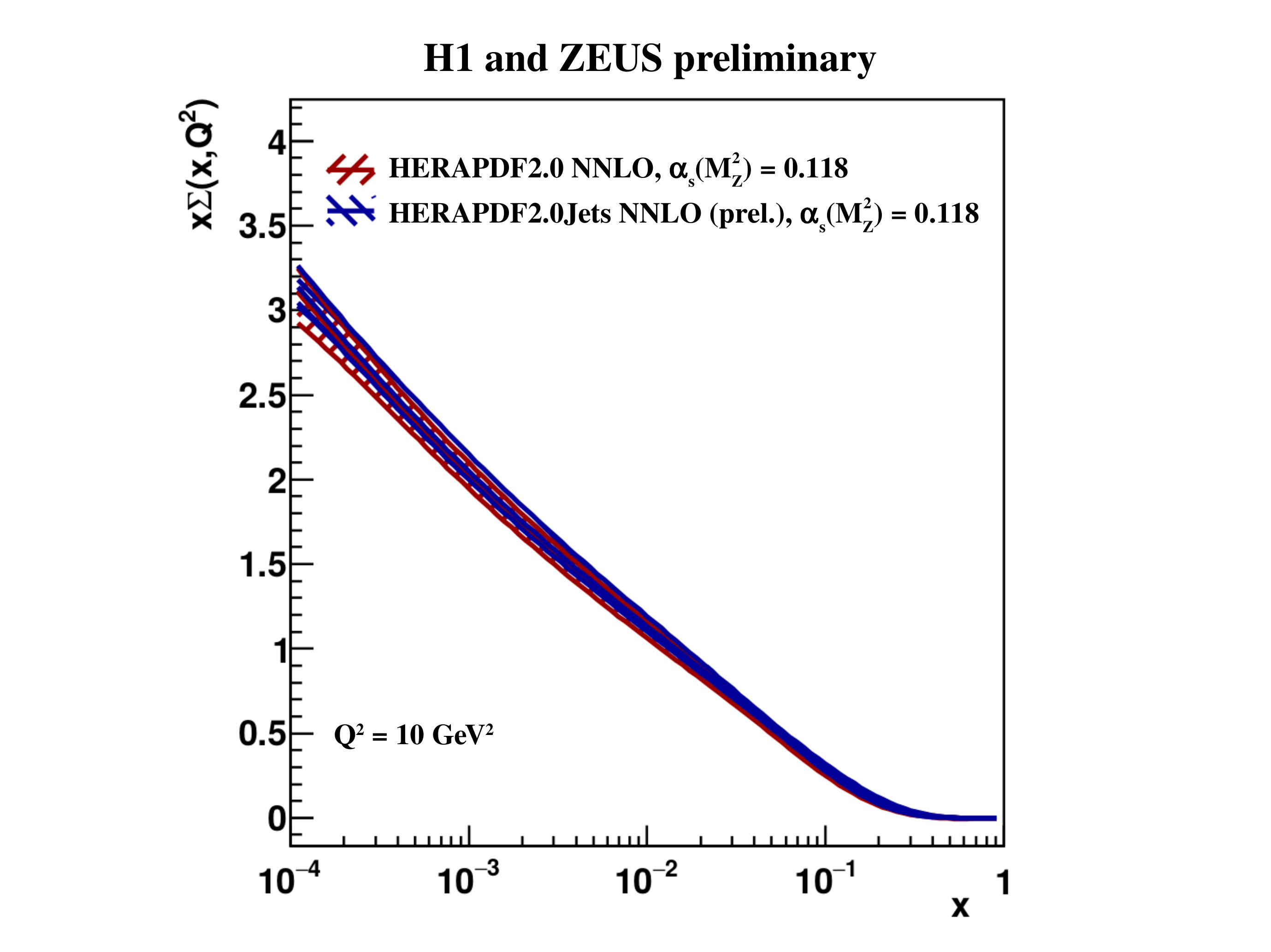}}
  \put (0.6,4.7) {a)}
  \put (5.6,4.7) {b)}
  \put (0.6,0.5) {c)}
  \put (5.6,0.5) {d)}
  \end{picture}
\vspace{-0.5cm} 
\caption { 
Comparison of the parton distribution functions 
a) $xu_v$, b) $xd_v$, c) $xg$ and d) $x\Sigma=x(\bar{U}+\bar{D})$ of 
HERAPDF2.0Jets NNLO (prel.) 
and HERAPDF2.0 NNLO based
on inclusive data only, both with fixed $\asmz = 0.118$,
at the scale $Q^{2} = 10\,$GeV$^{2}$.
The total uncertainties are shown as differently hatched bands.
}

\label{fig:as0-118vsherapdf2}
\end{figure}
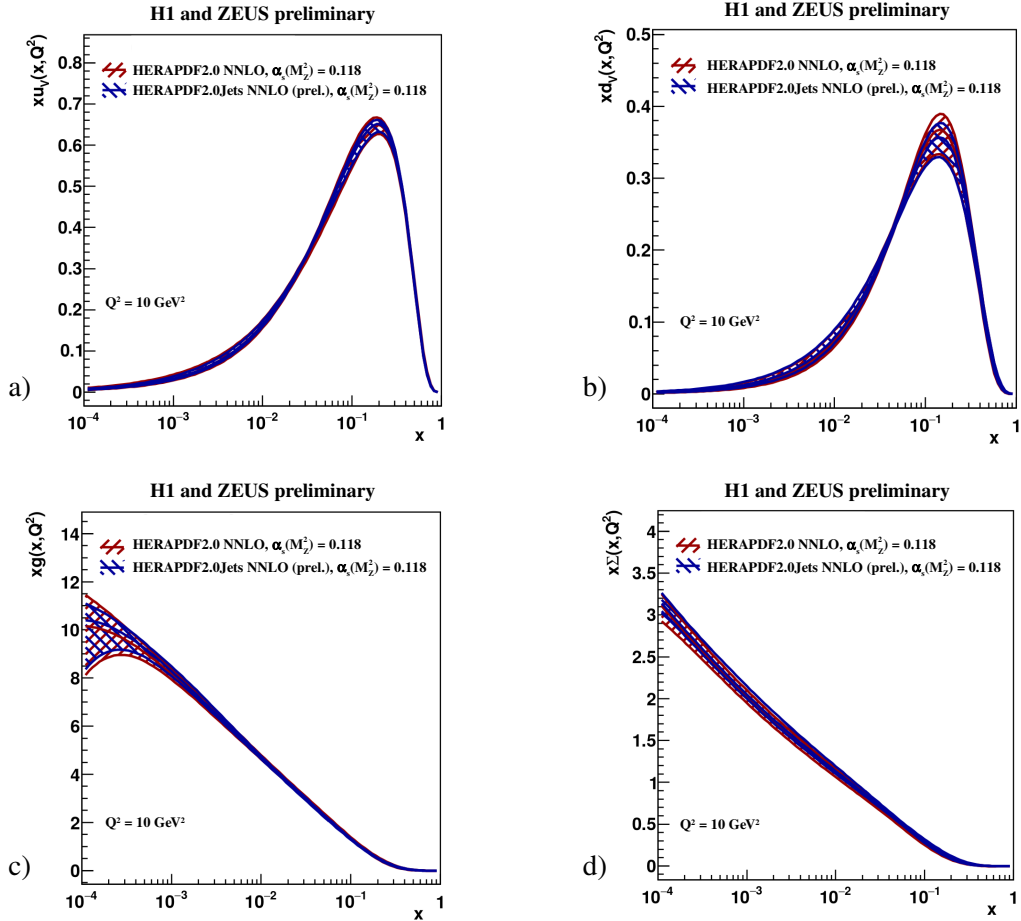

The PDFs resulting from the HERAPDF2.0Jets NNLO (prel.)  fit with
fixed $\asmz = 0.115$ and $\asmz = 0.118$ are shown and compared in
in Fig.~\ref{fig:as0-115vsas0-118} at a scale of
$Q^2=10$\,GeV$^2$. Here, total uncertainties are shown, including 
 experimental, model and parameterisation uncertainties as 
well as additional hadronisation uncertainties on the jet data. 
The former value of $\asmz$ is chosen because it is the preferred value of 
these data and the latter value is chosen because it is the PDG value, 
it also allows direct comparison to the published  
  PDFs of HERAPDF2.0 NNLO based on inclusive data only. 
These PDFs are very similar as shown in Fig.~\ref{fig:as0-118vsherapdf2}, 
indicating that the jet data do not change PDF 
shapes for fixed $\asmz$, but they have impact on the extracted value of 
$\asmz$, when it is allowed to be free. 
The comparison of the new HERAPDF2.0Jets NNLO (prel.) fits
with differing values of $\asmz$ shows a  
significant difference in the gluon distributions, as expected given the 
correlation between the gluon PDF shape and the value of $\asmz$.

\section{Summary}

The HERA data set on inclusive $ep$ scattering  as introduced by the 
ZEUS and H1 collaborations, 
together with selected data on jet production, published separately by 
the two collaborations, were used as input to NNLO fits
called HERAPDF2.0Jets NNLO (prel.). They complete the HERAPDF2.0 family.
A fit with free $\asmz$ gave
$\asmz = 0.1150 \pm 0.0008{\rm (exp)} ^{+0.0002}_{-0.0005} {\rm (mo-} $
${\rm del/parameterisation)} \pm 0.0006{\rm (hadronisation)}~~ \pm 0.0027 {\rm (scale)}$.
A preliminary set of PDFs with a full analysis of uncertainties was
obtained from a HERAPDF2.0Jets NNLO (prel.) fit with fixed $\asmz = 0.115$.
These PDFs were compared to PDFs from a similar fit with fixed $\asmz = 0.118$
and the PDFs from HERAPDF2.0 NNLO based on inclusive data only. All these PDFs
are very similar.


\begin{thebibliography}{99}
\bibitem{HERAPDF20} H.~Abramowicz {\it et al.} [ZEUS and H1 Collaboration], Eur. Phys. J. C {\bf 75}, 580 (2015), [arXiV:1506.06042].
\bibitem{h1lowq2newjets} V.~Andreev {\it et al.} [H1 Collaboration], Eur. Phys. J. C {\bf 77}, 215 (2017), [arXiV:1611.03421].
\bibitem{h1zeusprelim} H1prelim-19-041, ZEUS-prel-19-001, www-h1.desy.de/publications/htmlsplit/H1prelim-19-041.long.html
\bibitem{PDG18} M.~Tanabashi {\it et al.} [Particle Data Group], Phys. Rev.D {\bf 98}, 030001 (2018)
\bibitem{nnlojet} T.~Gehrmann {\it et al.}, arXiv:1801.06415
\bibitem{applfast} V.~Andreev {\it et al.} [H1 Collaboration], Eur. Phys. J. C {\bf 77}, 791 (2017)
\end{thebibliography}
\end{document}